\documentclass[onecolumn,secnumarabic,amsmath,amssymb,balancelastpage,nofootinbib]{article}

\usepackage[e]{esvect}
\usepackage{color}         
\usepackage{graphics}      
\usepackage{graphicx}      
\usepackage{epsf}          
\usepackage{bm}            

\usepackage{natbib}
\usepackage{amssymb}
\usepackage{amsmath, esint}
\usepackage{mathrsfs}
\usepackage{framed}
\usepackage{bigints} 
\usepackage{enumitem}
\usepackage{pifont}
\usepackage[capbesideposition={left,center},facing=yes,capbesidewidth=8cm,capbesidesep=quad]{floatrow} 
\usepackage{setspace} 

\usepackage[none]{hyphenat} 

\usepackage[colorlinks=true]{hyperref}  

\definecolor{darkred}{rgb}{0.6,0,0}
\definecolor{darkgreen}{rgb}{0,0.5,0}
\definecolor{darkblue}{rgb}{0,0,0.6}
\hypersetup{ colorlinks,
linkcolor=darkblue,
filecolor=darkgreen,
urlcolor=darkgreen,
citecolor=darkred }

\newcommand{\del}[0]{\ensuremath{\vec{\nabla}}}
\newcommand{\sigmaT}[0]{\ensuremath{\overset{\text{\tiny$\leftrightarrow$}}{\sigma}}}
\newcommand{\tensor}[1]{\ensuremath{\overset{\text{\tiny$\leftrightarrow$}}{#1}}}

\begin{document}

\sloppy 

\bibliographystyle{authordate1}


\title{\vspace*{-35 pt}\Huge{Forces on Fields}}
\author{Charles T. Sebens\\University of California, San Diego\vspace*{6 pt}\\Forthcoming in \textit{Studies in History and}\\\textit{Philosophy of Modern Physics}}
\date{February 9, 2018\\ arXiv v.3}

\maketitle
\begin{abstract}

In electromagnetism, as in Newton's mechanics, action is always equal to reaction.  The force from the electromagnetic field on matter is balanced by an equal and opposite force from matter on the field.  We generally speak only of forces exerted by the field, not forces exerted upon the field.  But, we should not be hesitant to speak of forces acting on the field.  The electromagnetic field closely resembles a relativistic fluid and responds to forces in the same way.  Analyzing this analogy sheds light on the inertial role played by the field's mass, the status of Maxwell's stress tensor, and the nature of the electromagnetic field.

\end{abstract}

\section{Introduction}

Newton's third law states that whenever one body exerts a force on a second, the second body exerts an equal and opposite force on the first.  The electromagnetic field exerts forces on matter via the Lorentz force law.  I will argue that matter exerts equal and opposite forces on the field.

Talk of forces on fields is generally resisted as fields seem too insubstantial to be acted upon by forces.  It would be hard to understand how fields could feel forces if they had neither masses nor accelerations.  Fortunately, fields have both.  Fields respond to forces in much the same way that matter does.

Few authors explicitly reject the idea that matter exerts forces on the electromagnetic field.  Instead, the rejection is implied by conspicuous omission.  In deriving and discussing the conservation of momentum, one speaks freely of the \emph{force} on matter but only of the \emph{rate of change of the momentum} of the electromagnetic field (e.g., \citealp{cullwick1952}; \citealp[section 8.2]{griffiths}; \citealp[section 4.9]{rohrlich}).

My primary goal in this article is to argue that Newton's third law holds in the special relativistic theory of electromagnetism because the force from the electromagnetic field on matter is balanced by an equal and opposite force from matter on the field.  I show that the field experiences forces by giving a force law for the electromagnetic field using hydrodynamic equations which describe the flow of the field's mass (originally studied by \citealp{poincare1900}).  In the course of this analysis I clarify the inertial role played by the field's mass---it quantifies the resistance the field itself has to being accelerated.   I also point out that Maxwell's stress tensor is in fact a momentum flux density tensor, not---as its title would suggest---a stress tensor, and give the true stress tensor for the electromagnetic field.  Finally, I explore the extent of the resemblance between the electromagnetic field and a relativistic fluid, asking (i) whether we can replace Maxwell's equations with fluid equations, (ii) if it is possible to understand the classical electromagnetic field as composed of photons, and (iii) how we can attribute proper mass to the field.

\section{Apparent Violation of the Third Law}\label{apparentviolation}

\begin{figure}[htb]\centering
\includegraphics[width=10 cm]{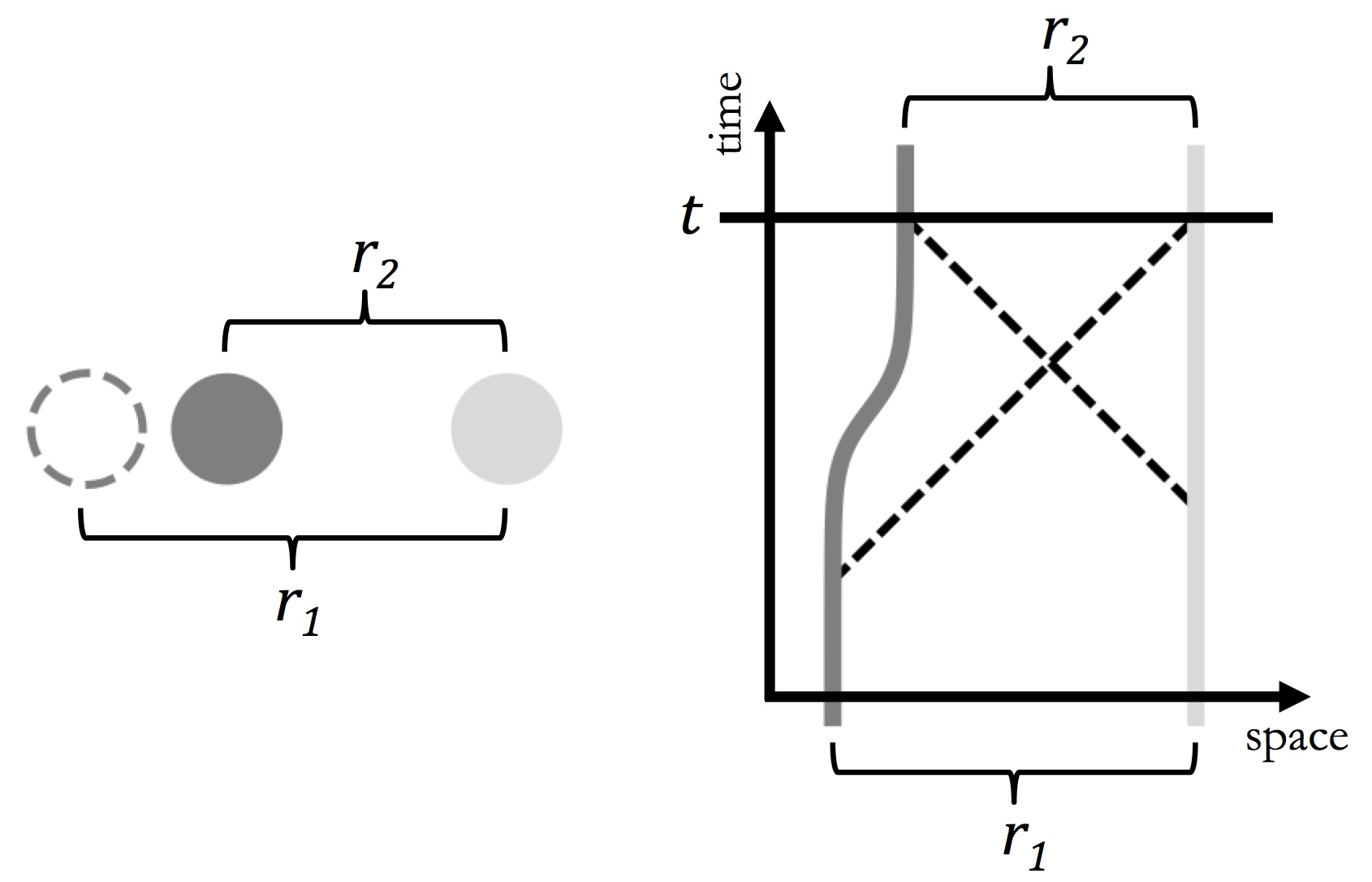}
\caption{The two gray lines represent spacetime trajectories of charged particles.  The dotted lines indicate which point one must examine on each particle's spacetime trajectory to calculate the force on the other at $t$---taking into account the light-speed delay on interactions.}
\label{particlepaths}
\end{figure}

If one takes charged particles to exert electromagnetic forces directly upon one another at a distance, violations of Newton's third law are easy to generate.  Consider the following case (\citealp[section 5.2]{lange}):  There are two particles of equal charge initially held in place (at rest) and separated by a distance $r_1$.  Then, one particle is quickly moved directly towards the other as depicted in figure \ref{particlepaths} so that at time $t$ the distance between the two particles is $r_2$.  Because there is a light-speed delay in the way charged particles interact with one another, the force that each particle feels from the other at $t$ cannot be calculated just by looking at what's going on at $t$.  The force on the stationary particle at $t$ is calculated by looking at the state of the particle that moved at the time when a light-speed signal from that particle would just reach the stationary particle at $t$.  At this earlier time, the particle was a distance $r_1$ from where the stationary particle is at $t$.  The general law describing how the force on one charge depends on the state of another at an earlier time is complex,\footnote{The law giving the force that one charged particle exerts on another is calculated from the retarded Li\'{e}nard-Wiechert potentials (\citealp[chapter 10]{griffiths}; \citealp[section 21-1]{feynman2}; \citealp[pg. 30]{lange}; \citealp[section 2]{earman2011}).  Newton's third law is violated in this sort of case because we are calculating forces directly between particles, not because of the particular choice to use retarded potentials in order to do so.  If advanced potentials were used instead, a similar violation would arise if the swerve were placed in the future instead of the past.  If half-retarded half-advanced potentials were used to calculate the forces between particles, either swerve would be sufficient to generate a violation.} but in this simple case where both particles are at rest at the relevant times, the repulsive force that the stationary particle feels at $t$ has magnitude $\frac{q^2}{r_1^2}$.  Similarly, the force on the particle that moved is calculated by looking at the state of the stationary particle at a time when the stationary particle was at a distance $r_2$ from where the particle that moved is at $t$.  The repulsive force the particle that moved feels at $t$ has magnitude $\frac{q^2}{r_2^2}$, opposite but not equal the force on the stationary particle.

As a second example \citep[section 8.2.1]{griffiths}, imagine two particles of equal charge, both equidistant from the origin and approaching at the same speed.  Particle 1 approaches along the $x$-axis from positive infinity and particle 2 along the $y$-axis.  Both are guided so that they unerringly follow their straight paths at constant speed.  In this case the electric forces on the two particles are equal and opposite but the magnetic forces are equal in magnitude but not opposite in direction.  The magnetic force on particle 1 is in the $y$-direction whereas the magnetic force on 2 is in the $x$-direction.

According to Griffiths, we should be troubled by this violation because ``...the proof of conservation of momentum rests on the cancellation of internal forces, which follows from the third law.  When you tamper with the third law, you are placing the conservation of momentum in jeopardy, and there is no principle in physics more sacred than \emph{that}.''  Griffiths then immediately neutralizes the threat, writing that ``Momentum conservation is rescued in electrodynamics by the realization that the fields themselves carry momentum.''  \citet[sections 26-2 and 27-6]{feynman2} respond to apparent violations of the third law in a similar manner.  They write that they will leave it to the reader to worry about whether action is equal to reaction, but point out that momentum is conserved---provided that the field momentum is included---and seem satisfied with this resolution of the puzzle.

I believe these responses capture the general attitude of physicists to the apparent violation of Newton's third law and they are correct as far as they go.  However, by shifting the focus to conservation of momentum they leave the question of whether Newton's third law holds unanswered.  Since conservation of momentum has been upheld and the status of Newton's third law remains uncertain, one might reasonably conclude that conservation of momentum is the deeper principle.  This common attitude appears in the \citet{wikipedia} article on Newton's laws of motion: ``Newton used the third law to derive the law of conservation of momentum; from a deeper perspective, however, conservation of momentum is the more fundamental idea (derived via Noether's theorem from Galilean invariance), and holds in cases where Newton's third law appears to fail, for instance when force fields as well as particles carry momentum, and in quantum mechanics.''  \citet[pg. 163]{lange} gives a more definitive rejection of the third law as a footnote to his discussion of conservation of energy and momentum, ``However, Newton's third Law (`Every action is accompanied by an equal and opposite reaction') is still violated, even if fields are real.  Bodies do not exert forces on fields; bodies alone feel forces.  Newton's third law was thus abandoned before relativity theory came on the scene.''\footnote{According to \citet[pg. 16]{frisch2016}, \citet{ritz1908} made a similar point while defending a version of electromagnetism without an electromagnetic field (in which charged particles act directly upon one another) and criticizing versions of the theory that include field or aether: ``[Ritz] also notes that a theory presupposing an aether does not obey the equality of action and reaction, since the particle does not react back when the aether acts on a particle.''}

Another possible reaction to our quandary is to view the third law as immediately saved by the fact that momentum is conserved.  If force is simply the rate of change of momentum, then the fact that the amount of momentum in the field is changing is sufficient to demonstrate that forces act on the field (presumably from matter as it is the only other actor on the scene).  Because momentum is conserved, changes in momentum must cancel and thus forces must balance---Newton's third law is preserved.  I think it is ultimately correct that the third law is saved by the fact that forces act on fields.  However, I find this quick version of the argument unsatisfactory.  One reason for dissatisfaction is that although the presence of forces on fields is suggested, a mathematical account of how forces act on fields is absent.  Another problem with this quick argument is that it begs the question against someone who thinks that the conservation of momentum is a deeper principle than Newton's third law and may hold in cases where Newton's third law does not, as this argument makes obedience of the third law an immediate consequence of the conservation of momentum.

Some readers might balk at the idea that forces could act upon the electromagnetic field because they think that the field is merely a useful tool, not a real thing.  If the field isn't real, it's hard to see how either Newton's third law or the conservation of momentum could hold (though some clever maneuvers have been made to save Newton's third law and conservation of momentum in field-less versions of electromagnetism; see \citealp{wheeler1949}; \citealp[chapter 5]{lange}; \citealp[section 4.2]{Lazarovici2017}).  Over the years, much has been said in favor of, and in opposition to, taking the electromagnetic field to be real.  For my purposes here, I would like to avoid entering this debate by simply assuming a certain resolution---that the field is real---and addressing the status of the third law given this assumption.  Once complete, one might take the story presented here to provide new reasons for believing the field to be real.  But, I will not explicitly draw them out as this debate is not my focus.

In what follows I will give a more thorough defense of the idea that forces act on fields and explain how they do so.  I take the mark of a force to be the obedience of something like Newton's second law, $\vec{F}=m\vec{a}$.  However, we continue to speak in terms of forces despite certain modifications to that simple equation.  In particular, the second law can be extended to special relativitistic continuum mechanics.  What further modifications the law can sustain while still counting as a force law is more a choice of convention than a question of deep metaphysics.  Perhaps the best convention is to be liberal about such modifications and to say that wherever there is change in momentum there is force.  Fortunately, we need not judge such modifications in order to determine whether the electromagnetic field experiences forces.  There are two laws of relativistic continuum mechanics which might deserve to be called force laws and both are obeyed by the electromagnetic field.  Establishing this requires making sense of an $m\vec{a}$ type reaction of the field to the forces it experiences.  We turn to this problem in the next two sections.

\section{The Eulerian Perspective}\label{fmfv}

In extending Newton's second law from the case of discrete bodies to continuum mechanics we face a choice as to how to describe the physics.  There are two equations that might deserve to be called ``the force law,'' an Eulerian and a Lagrangian equation of motion.  In this section, I give an Eulerian force law for the electromagnetic field \eqref{momentumconservationfield} by first explaining how one can attribute mass and velocity to the field.  The Eulerian force law makes clear what force matter exerts on the field but obscures the nature of field-on-field forces. In the next section, I give a Lagrangian force law \eqref{finalforcelaw} which clarifies these forces.

Nonrelativistically, bodies respond to forces according to $\vec{F}=m\vec{a}$ where the mass $m$ quantifies the resistance to acceleration.  Relativistically, bodies respond to forces by
\begin{equation}
\vec{F} = \frac{d\vec{p}}{dt}=\frac{d}{dt}\left(m_r\vec{v}\right)\ ,
\label{relativisticforce}
\end{equation}
where $\vec{p}=m_r\vec{v}$ is the body's momentum and $m_r$ is the velocity-dependent relativistic mass of the body, related to its proper mass by $m_r=\gamma m_0$ with $\gamma=\left(1-\frac{v^2}{c^2}\right)^{-\frac{1}{2}}$.  (In this article I will use the word ``mass'' without prefix to mean relativistic---not proper---mass, as it is relativistic mass which is most central to the analysis here.)  Because the time derivative of $m_r$ is taken in \eqref{relativisticforce}, mass no longer acts as a simple constant of proportionality between force and acceleration.  To clarify the new way in which $m_r$ quantifies resistance to acceleration in relativistic particle mechanics, the force law can be rewritten as\footnote{This alternative form is arrived at by expanding the derivative of $\gamma$ in \eqref{relativisticforce}; see \citet[pg. 49]{landaulifshitzfields} for the application to electromagnetic force.}
\begin{equation}
\vec{F}-\frac{\vec{v}}{c^2}\left( \vec{v} \cdot \vec{F}  \right)= m_r\vec{a}\ .
\label{relativisticforcealternative}
\end{equation}

If we are interested in the force per unit volume on a continuous distribution of mass instead of the force on a discrete body, \eqref{relativisticforce} can be replaced by
\begin{equation}
\vec{f}=\frac{\partial}{\partial t}\left(\rho_m \vec{v}_m\right)-\vec{\nabla}\cdot \sigmaT_{\!m}\ ,
\label{momentumconservationfluid}
\end{equation}
an Eulerian force law familiar from fluid mechanics.  Here $\vec{f}$ is the force per unit volume on the fluid, $\rho_m$ is the fluid's relativistic mass density, $\vec{v}_m$ is a velocity field describing the flow of this mass, $\rho_m \vec{v}_m$ is the momentum density, and $\sigmaT_{\!m}$ is the momentum flux density tensor for the fluid.\footnote{The tensor $\sigmaT_{\!m}$ used here is equal to minus the momentum flux density tensor as it is usually defined (see the tensor $\tensor{\Pi}$ in \citealp[equation 23.16]{mihalas}; \citealp[pg. 47]{landaulifshitzfluids}).}  All of these quantities are functions of space and time.  The subscript $m$ denotes properties of matter (as opposed to $f$ for field). If the source of the force on matter is the electromagnetic field, $\vec{f}$ is the Lorentz force per unit volume,
\begin{equation}
\vec{f} = \rho^q_m \left( \vec{E} + \frac{1}{c}\vec{v}^{\,q}_m \times\vec{B} \right)\ ,
\label{lorentzforce}
\end{equation}
where $\rho^q_m$ is the charge density and $\vec{v}^{\,q}_m$ is the velocity field describing the flow of charge.\footnote{For simplicity, I assume that the matter has a charge density but neither a magnetic nor an electric dipole moment density.  Such properties would complicate the interaction between matter and field, e.g., in \eqref{lorentzforce}.}  The superscript $q$ denotes that these quantities describe the flow of charge as opposed to mass.  Section \ref{propermassdensitysec} explains why $\vec{v}^{\,q}_m\neq\vec{v}_m$.

The Eulerian force law for a continuous distribution of mass can alternatively be expressed in integral form by integrating \eqref{momentumconservationfluid} over an arbitrary volume $V$ and applying the divergence theorem to $\vec{\nabla}\cdot \sigmaT_{\!m}$,
\begin{equation}
\iiint \frac{\partial}{\partial t}(\rho_m \vec{v}_m) dV =\iiint \vec{f}\: dV+ \oiint \sigmaT_{\!m}\cdot \hat{n}\: dA\ ,
\label{integralformmomentumconservationfluid}
\end{equation}
where $\hat{n}$ is a unit vector normal to the surface of the volume.  This equation equates the rate at which the momentum of matter in a fixed volume changes over time with the force exerted on the matter in that volume plus the flux of momentum into that volume from matter outside.

Although I speak of matter as a fluid, nothing I say depends on whether it is solid, liquid, or gas (or some combination).  The equations I am appealing to---such as \eqref{momentumconservationfluid} and \eqref{integralformmomentumconservationfluid}---are equations of continuum mechanics, valid for solids as well as fluids.  Use of this continuum-level description is neutral as to whether the matter under discussion is ultimately continuous or particulate.  I refer to the matter as a fluid and imagine it as ultimately composed of charged particles because I think having such a concrete picture in mind is helpful in developing a physical understanding of the mathematics.  I do not mean to be making any substantive assumptions about the kind of matter that is interacting with the field.

Momentum is conserved in electromagnetism because change in the momentum of matter, as expressed in \eqref{momentumconservationfluid}, is balanced by a compensating change in the momentum of the electromagnetic field,
\begin{equation}
-\vec{f}=\frac{\partial}{\partial t}\left(\frac{\vec{S}}{c^2}\right)-\vec{\nabla}\cdot \sigmaT_{\!f}\ .
\label{momentumconservation}
\end{equation}
This equation can be derived from Maxwell's equations and the Lorentz force law.  The momentum density of the electromagnetic field is $\frac{\vec{S}}{c^2}$, where $\vec{S}$ is the Poynting vector which gives the energy flux density,
\begin{equation}
\vec{S}= \frac{c}{4\pi} \vec{E} \times \vec{B}\ .
\label{poyntingvector}
\end{equation}
The (symmetric) tensor $\sigmaT_{\!f}$ in \eqref{momentumconservation} is the momentum flux density tensor of the electromagnetic field (also known as the ``Maxwell stress tensor,'' though I'll explain why I dislike that name later) which can be expressed in terms of the electric and magnetic fields as,
\begin{equation}
\sigmaT_{\!f}=\frac{1}{4\pi}\vec{E}\otimes\vec{E}+\frac{1}{4\pi}\vec{B}\otimes\vec{B}-\frac{1}{8\pi}\left(E^2+B^2\right)\tensor{I} \ ,
\label{maxwellstresstensor}
\end{equation}
where $\otimes$ is the tensor product and $\tensor{I}$ is the identity tensor.\footnote{In this paper I use the standard expressions for the energy density, momentum density, and momentum flux density tensor of the electromagnetic field.  These quantities together form the field's four-dimensional symmetric stress-energy tensor \citep[section 12.10]{jackson}.}  Here and throughout I use cgs units.

The equations which quantify the rate at which matter and field momenta change, \eqref{momentumconservationfluid} and \eqref{momentumconservation}, are quite similar.  This similarity suggests taking $-\vec{f}$ to be a force exerted by matter on the field, equal and opposite to that of the field on matter.  However, the momentum of the electromagnetic field is an exotic thing and its change does not obviously resemble the ordinary response of a body to a force.  The momentum density of the electromagnetic field is $\frac{\vec{S}}{c^2}$, not $\rho_m \vec{v}_m$.  To show that the field responds to forces in just the same way that matter does, we must attribute mass and velocity to the field.

The fact that the electromagnetic field has mass can be seen as a result of the special relativistic equivalence of mass and energy (\citealp[pg. 204]{einstein1906}).  The field's relativistic mass density is equal to the energy density over $c^2$ (by $\mathcal{E}=m_r c^2$),
\begin{equation}
\rho_f=\frac{1}{8 \pi c^2}\left(E^2+B^2\right)\ .
\label{massdensityfield}
\end{equation}
In general, the (relativistic) mass of matter will vary over time as it exchanges energy with the electromagnetic field.  Attributing the above mass to the electromagnetic field ensures that the total mass of matter and field is conserved and that the center of mass for a closed system moves inertially \citep{einstein1906}.  These considerations provide strong motivation for saying that the field has mass. But, one might wonder: Does the field act like it truly has \emph{inertial} mass, the kind of mass that quantifies resistance to acceleration?  It has often been noted that the field around or inside a body can contribute to the inertial mass of that body (e.g., consider accelerating a box of radiation or a spherical charge and its surrounding field; see \citealp[pg. 51--52]{whittaker2}; \citealp[section 5]{griffithsletter}).  But, this is an indirect way to get at the inertial role played by the mass of the field.  The mass of the field does not merely make it more difficult to accelerate bodies.  We will see that the mass of the field quantifies the resistance to acceleration \emph{of the field itself}, just as the mass of a fluid quantifies its resistance to acceleration.

In addition to field mass, we must also make sense of field velocity if we are to understand forces on fields analogously to forces on matter.  The field velocity can be found by analyzing the flow of energy.  The conservation of energy for the electromagnetic field is expressed by Poynting's theorem (which, like \eqref{momentumconservation}, is derivable from Maxwell's equations and the Lorentz force law),
\begin{align}
\frac{\partial}{\partial t}\left[\frac{1}{8 \pi}\left(E^2+B^2\right)\right]+\vec{\nabla}\cdot \vec{S}&=-\vec{f}\cdot\vec{v}^{\,q}_m
\nonumber
\\
&=-\left[\frac{\partial}{\partial t}\left(\rho_m c^2\right) + \vec{\nabla}\cdot \left(\rho_m \vec{v}_m c^2\right)\right]\ .
\label{energycons}
\end{align}
Here I've equated the change in energy of the field with the change in energy of the matter.  The rate of energy transfer between fields and matter, $\vec{f}\cdot\vec{v}^{\,q}_m$, is equal to $\vec{J}\cdot\vec{E}$ where $\vec{J}=\rho_m^q\vec{v}^{\,q}_m$ is the current density.\footnote{Being careful, velocity fields like $\vec{v}^{\,q}_m$ only give a coarse-grained description of the velocities of the particles which compose the fluid.  If we assume all of the particles have equal charge, $\vec{v}^{\,q}_m$ assigns to each point in space at each time the average velocity of particles in a very small volume around that point over a very short period of time (\citealp[section 2.2]{chapman}, \citealp[section 1.2]{batchelor}).   There will be particles with velocities unequal to the mean velocity and they will experience different Lorentz forces.  However, because the Lorentz force depends linearly on velocity, the average velocity can be used to correctly calculate the force per unit volume on the left-hand side of \eqref{momentumconservationfluid} via \eqref{lorentzforce}.  The rate of energy transfer is also linearly dependent on the velocity (since it is proportional to the dot product of the velocity and the electric field), so the average velocity can be used on the right hand side of the first line of \eqref{energycons} as well.\label{average}}  In the second line of \eqref{energycons}, $\rho_m c^2$ is the energy density of matter and $\rho_m \vec{v}_m c^2$ is the energy flux density.  Divide each term in \eqref{energycons} by $c^2$ and it expresses the conservation of mass.  The first term becomes $\frac{\partial \rho_f}{\partial t}$ by \eqref{massdensityfield}.  When integrated over a volume, the second term, $\vec{\nabla}\cdot \left(\frac{\vec{S}}{c^2}\right)$, gives the rate at which the field's mass leaves the volume.  If we treated the field as having a velocity, that term would be $\vec{\nabla}\cdot (\rho_f \vec{v}_f)$ and the equation would become,
\begin{align}
\frac{\partial \rho_f}{\partial t}+\vec{\nabla}\cdot \left(\rho_f \vec{v}_f\right)&=\frac{-\vec{f}\cdot\vec{v}^{\,q}_m}{c^2}
\nonumber
\\
&= -\left[\frac{\partial \rho_m}{\partial t} + \vec{\nabla}\cdot (\rho_m \vec{v}_m)\right]\ .
\label{emcontinuity}
\end{align}
This suggests taking the field velocity to be
\begin{equation}
\vec{v}_f=\frac{\vec{S}}{\rho_f c^2}\ ,
\label{fieldvelocity}
\end{equation}
equal to the field momentum density divided by the mass density or, equivalently, equal to the energy flux density divided by the energy density.\footnote{There are other possible definitions of field velocity that agree on the rate at which field mass leaves any closed volume.  However, such alternative definitions would not retain the appropriate relations between velocity, energy flux density, and momentum density.  Still, one could cast doubt on this definition of field velocity by challenging the standard expressions for energy flux density and momentum density---or even energy density (see \citealp[section 27-4]{feynman2}; \citealp[pg. 347, footnote 1]{griffiths}; \citealp{lange2001, lange}; \citealp[sections 31-33]{landaulifshitzfields}; \citealp[sections 6.7 and 12.10]{jackson}).}  This velocity is not often discussed, but when it is it goes by the name of the ``velocity of energy transport''---though we could just as well call it the ``velocity of (relativistic) mass transport'' (\citealp{poincare1900}, \citealp[section 14.2.1]{bornwolf}; \citealp[section 12.6.2]{holland}; \citealp[box 8.3]{lange}).  The magnitude of the velocity is maximized at $c$ when $\vec{E}$ is perpendicular to $\vec{B}$ and $|\vec{E}|=|\vec{B}|$.

Using the mass and velocity just introduced, the conservation of momentum equation for the field \eqref{momentumconservation} looks just like the Eulerian force law for matter \eqref{momentumconservationfluid},
\begin{equation}
-\vec{f}=\frac{\partial}{\partial t}\left(\rho_f \vec{v}_f\right)-\vec{\nabla}\cdot \sigmaT_{\!f}\ .
\label{momentumconservationfield}
\end{equation}
Putting \eqref{momentumconservationfluid} and \eqref{momentumconservationfield} together, we see that the matter and the field interact like two overlapping fluids, exerting equal and opposite forces on one another via
\begin{align}
\frac{\partial}{\partial t}\left(\rho_f \vec{v}_f\right)-\vec{\nabla}\cdot \sigmaT_{\!f}&=-\vec{f}
\nonumber
\\
&=-\left[\frac{\partial}{\partial t}\left(\rho_m \vec{v}_m\right)-\vec{\nabla}\cdot \sigmaT_{\!m}\right]\ ,
\label{together1}
\end{align}
and exchanging mass with one another by \eqref{emcontinuity}.  The interaction between matter and field is entirely local.  The force density $\vec{f}$ on matter at a point in space is balanced exactly by a force density $-\vec{f}$ on the field.

\citet{poincare1900} proposed that we treat the energy of the electromagnetic field as a ``fictitious fluid,'' with mass and velocity essentially as described above, in order to recover Newton's third law.\footnote{See \citet{weinstein2012} for a helpful translation of Poincar\'{e}'s equations into modern notation.}  Poincar\'{e} sometimes talks as if matter can exert forces on the electromagnetic field, writing that the forces on a volume must be balanced by the inertial forces of the matter and the inertial forces of the fictitious fluid.  He explains the way in which the action on matter is matched by an equal and opposite action on the field as follows: ``...since the electromagnetic energy behaves as a fluid which has inertia, we must conclude that, if any sort of device produces electromagnetic energy and radiates it in a particular direction, that device must \emph{recoil} just as a cannon does when it fires a projectile.''  Poincar\'{e}'s nice story is marred by the fact that he considers this fluid and its associated mass to be fictitious.\footnote{\citet{stein2014} has argued that this failure to properly appreciate the true possession of inertial mass by the field was a key reason why Poincar\'{e} did not arrive at Einstein's special theory of relativity and symptomatic of his preference for systematizing existing knowledge as opposed to generating novel consequences.  Poincar\'{e} had hoped for the principle of action and reaction to be satisfied among matter alone and resisted the idea that the ether (or the electromagnetic field) is a real physical entity on a par with ordinary matter.}  In the 1900 paper, Poincar\'{e} wrote that the fluid should not be considered real because it can be created and destroyed.  That alone cannot be the problem.  I was created and I will be destroyed.  And yet, I exist.  Perhaps it is only unacceptable for mass to be created and destroyed.  However, from a modern (special relativistic) perspective this is commonplace---the relativistic mass of ordinary matter changes over time as well (as it exchanges energy with the field via \eqref{emcontinuity}).  If we consider field and matter together, there is no creation or destruction of mass (just transfer).  Poincar\'{e} has identified a way in which relativistic mass is unlike non-relativistic mass, but I don't believe he has given us any reason to think the field, its mass, or its velocity are not real.

In understanding the resemblance of field to fluid noted by Poincar\'{e} and explored in this section, it is helpful to contrast it with a more widely known analogy between hydrodynamics and electromagnetism utilized by \citet{maxwelllinesofforce} in ``On Faraday's Lines of Force.''  Faraday made great progress understanding electromagnetic phenomena in terms of electric and magnetic lines of force, precursors to the modern electric and magnetic fields.  Inquiring into ``the nature of the lines of magnetic force,'' Faraday at one point puts forward the hypothesis that ``physical lines of magnetic force are currents,'' seeing that these lines of force could potentially be understood as describing some kind of flow (\citealp[section 3269]{faraday3}; \citealp[pg. 243]{whittaker1}; \citealp[pg. 112]{darrigol}).  Using Faraday's lines of force, Thomson developed an analogy between electrostatics and heat flow---in which heat is taken to flow along electric lines of force---and an analogy between magnetism and hydrodynamics---in which an incompressible fluid is taken to flow along magnetic lines of force (\citealp[pg. 77--80]{harman1998}; \citealp[pg. 114--118, 128--132, 136]{darrigol}).  Building on the work of Faraday and Thomson, \citet{maxwelllinesofforce} introduced, as ``a purely imaginary substance,'' one incompressible fluid that flows along the electric lines of force and a second fluid that flows along magnetic lines of force (\citealp[pg. 242--244]{whittaker1}; \citealp[pg. 85--88]{harman1982}; \citealp[pg. 87--90]{harman1998}; \citealp[pg. 142--147]{darrigol}; \citealp[pg. 47--50]{lange}).  These heat and fluid analogies played an important role in the development of electromagnetic theory as they allowed Thomson and Maxwell to import mathematical tools and physical insight from better understood physical theories.  It should be clear that Maxwell's fluid analogy is quite different from Poincar\'{e}'s.  Poincar\'{e} treated the electric and magnetic fields together as a single compressible fluid which flows in the direction of energy flux.

\section{The Lagrangian Perspective}\label{forcelawforfield}

\begin{figure}[htb]\centering
\includegraphics[width=9 cm]{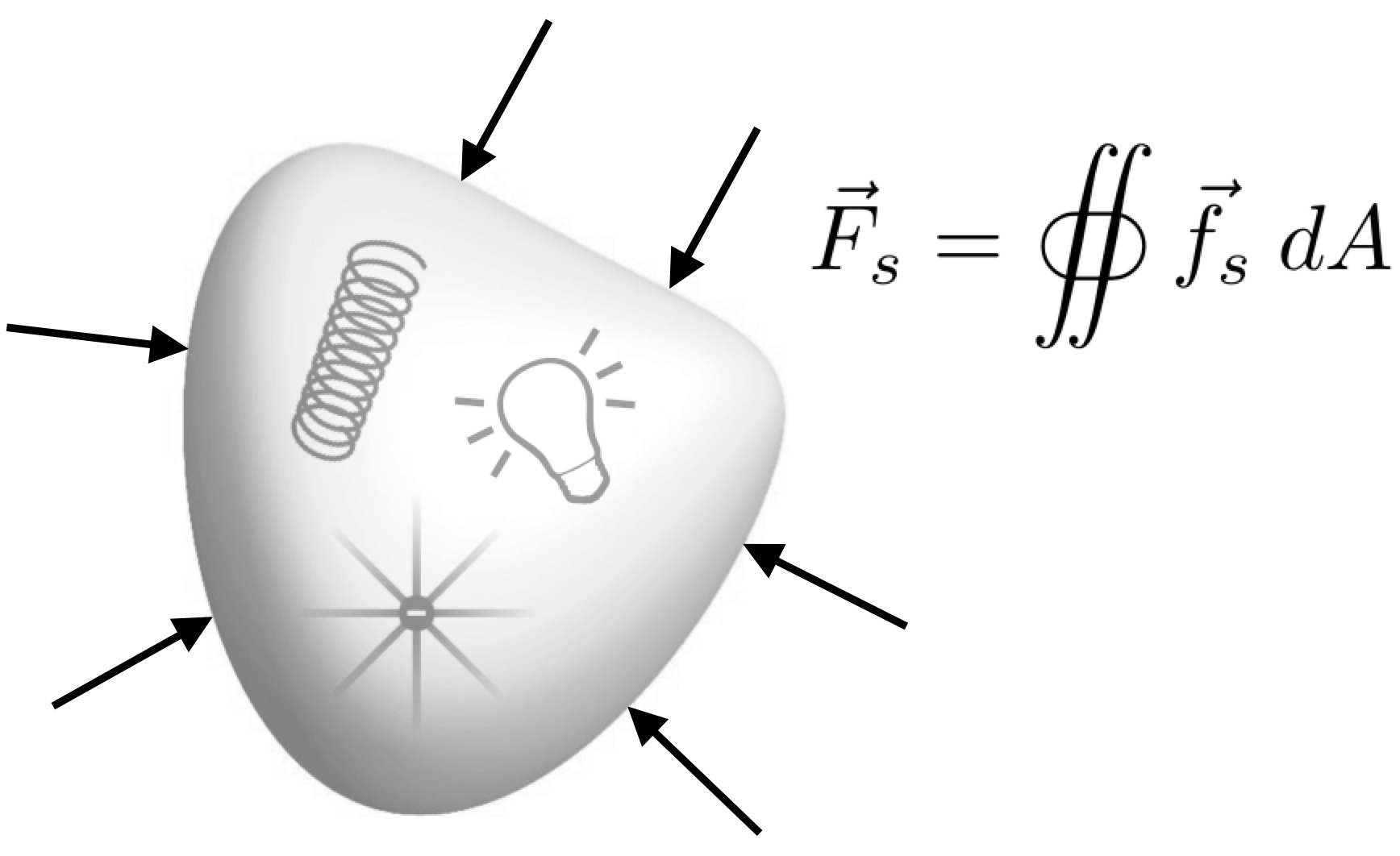}
\caption{This figure depicts the surface force $\vec{F}_s$ exerted by the field outside of a given volume containing varied charged particles and fields.}
\label{surfaceforces}
\end{figure}

Although the Eulerian force law \eqref{momentumconservationfield} captures the forces from matter on the field well, it does not make clear what forces the field exerts on itself.  To see the problem, consider integrating \eqref{momentumconservationfield} over an arbitrary volume $V$ to arrive at an integral form similar to \eqref{integralformmomentumconservationfluid},
\begin{equation}
\iiint \frac{\partial}{\partial t}(\rho_f \vec{v}_f) dV =\iiint \left(-\vec{f}\:\right)dV+ \oiint \sigmaT_{\!f}\cdot \hat{n}\: dA\ .
\label{integralform}
\end{equation}
One might incorrectly read this as saying that the reaction of the field in the volume is equal to the sum of a body force and a surface force where the body force has a density per unit volume of $\vec{f}_b=-\vec{f}$ which is integrated over the volume to give the total force from matter on the field and the surface force has a density per unit area with normal $\hat{n}$ of $\vec{f}_s= \sigmaT_{\!f}\cdot \hat{n}$ which is integrated over the surface to give the total force on the field inside the volume from the field outside the volume (see figure \ref{surfaceforces}).  \citet[pg. 261]{jackson} makes this misstep when he writes that if his equation for the conservation of momentum is correct, $\sigmaT_{\!f}\cdot\hat{n}$ gives ``the flow per unit area [with normal unit vector $\hat{n}$] of momentum across the surface $S$ into the volume $V$.  In other words, it is the force per unit area transmitted across the surface $S$ and acting upon the combined system of particles and fields inside $V$.''\footnote{I should note that the way Jackson phrases this point suggests that forces can act on fields, though he is not very explicit about this.}  This reading of \eqref{integralform} erroneously equates momentum flux and force.  The first term on the right side of \eqref{integralform} is indeed a body force on the field.  However, the second term is not a surface force giving the force from the field outside the volume (as would be suggested by the fact that $\sigmaT_{\!f}$ is called the stress tensor).  It is the net flux of (field) momentum into the volume.  The force on the field from matter is correct but the force from the field is not.  To properly understand this force, we must move from an Eulerian to a Lagrangian description.  Because the electromagnetic field obeys an Eulerian force law, it will also obey a Lagrangian force law.\footnote{Deriving the Lagrangian force law and thereby finding a mathematical expression for field-on-field forces requires an understanding of the field's velocity, not just its momentum.  This gives us further reason to attribute velocity to the field (in addition to momentum).}

The problem with reading the surface force off of \eqref{integralform} stems from the fact that the left side of the equation gives the change in the field momentum contained within a fixed volume, not the change in momentum of the field mass which happens to be at that moment contained within the volume.  The rate of change in momentum of that field mass is given by
\begin{equation}
\frac{D}{Dt} \iiint \rho_f \vec{v}_f \:dV\ ,
\label{emresponse}
\end{equation}
where $\frac{D}{Dt}$ is the material derivative which gives the rate of change of a quantity while following the flow (it is also known as the comoving or Lagrangian derivative).  Taking the material derivative inside the integral, \eqref{emresponse} becomes
\begin{equation}
=\iiint \left[ \frac{D}{Dt}\left(\rho_f \vec{v}_f\right) + \rho_f \vec{v}_f \left(\del \cdot \vec{v}_f\right) \right]dV \ ,
\end{equation}
where the second term accounts for the fact that we must consider the deformation of the volume when calculating the time evolution (see the discussion of the Reynolds transport theorem in \citealp[section 2.1]{mihalas}; \citealp[section 1.3.2]{slattery}; \citealp[sections 3.5.3 and 7.12.1]{belytschko}; \citealp[section 10]{gurtin}).  Using the product rule on the material derivative yields
\begin{equation}
=\iiint  \left[ \rho_f\frac{D\vec{v}_f}{Dt}+\vec{v}_f\frac{D\rho_f}{Dt}+ \rho_f \vec{v}_f \left(\del \cdot \vec{v}_f\right) \right]dV \ .
\label{stepxxxx}
\end{equation}
Expanding the material derivatives gives
\begin{equation}
=\iiint \left[ \rho_f\frac{\partial \vec{v}_f}{\partial t}+\rho_f\left(\vec{v}_f\cdot\del\right)\vec{v}_f+\vec{v}_f\frac{\partial \rho_f}{\partial t}+ \vec{v}_f \left( \vec{v}_f \cdot \del\right)\rho_f+ \rho_f \vec{v}_f \left(\del \cdot \vec{v}_f\right)  \right]dV \ .
\end{equation}
This simplifies to
\begin{equation}
=\iiint \left[\frac{\partial}{\partial t}\left(\rho_f\vec{v}_f\right)+\del\cdot\left(\rho_f\: \vec{v}_f\otimes\vec{v}_f\right)\right]dV\ .
\end{equation}
Using the conservation of momentum \eqref{momentumconservationfield} to expand the first term, this becomes
\begin{equation}
=\iiint \left[-\vec{f}+\del\cdot\left(\sigmaT_{\!f} + \rho_f\: \vec{v}_f\otimes\vec{v}_f\right)\right]dV\ .
\label{almostthere}
\end{equation}
Applying the divergence theorem, we arrive at
\begin{equation}
\frac{D}{Dt} \iiint \rho_f \vec{v}_f \:dV = \iiint  \left(-\vec{f}\:\right)dV + \oiint \left[\sigmaT_{\!f} + \rho_f\left(\vec{v}_f\otimes\vec{v}_f\right)\right]\cdot \hat{n}\: dA\ .
\label{finalforcelaw}
\end{equation}
This is the Lagrangian force law for the electromagnetic field.  The left-hand side is the response of the electromagnetic field in a region to forces acting on it.  The first term on the right-hand side is the force resulting from interaction with matter.  It is equal and opposite the Lorentz force of the fields on matter.  The second term on the right-hand side is a surface force from the electromagnetic field outside of the volume (figure \ref{surfaceforces}).

By parallel reasoning to the derivation of \eqref{finalforcelaw} (except that the force is opposite), the matter will satisfy
\begin{equation}
\frac{D}{Dt} \iiint \rho_m \vec{v}_m \:dV = \iiint  \vec{f}\:dV + \oiint \left[\sigmaT_{\!m} + \rho_m\left(\vec{v}_m\otimes\vec{v}_m\right)\right]\cdot \hat{n}\: dA\ .
\label{finalforcelawmatter}
\end{equation}
As is clear from comparing \eqref{finalforcelaw} and \eqref{finalforcelawmatter}, the mass of the electromagnetic field plays exactly the same inertial role in resisting acceleration as does the mass of a relativistic fluid.

In fluid mechanics, it is $\sigmaT_{\!m} + \rho_m\left(\vec{v}_m\otimes\vec{v}_m\right)$, not $\sigmaT_{\!m}$, which gets the name ``stress tensor'' because integrating it over a surface gives the force exerted by the matter on that surface (\citealp[equation 23.17]{mihalas}; \citealp[section 15]{landaulifshitzfluids}).  Similarly, we should call $\sigmaT_{\!f} + \rho_f\left(\vec{v}_f\otimes\vec{v}_f\right)$ the stress tensor for the electromagnetic field because integrating it over a surface gives the force exerted by the field on that surface.  However, that name is already taken by $\sigmaT_{\!f}$ itself, the Maxwell stress tensor.\footnote{For discussion of how Maxwell arrived at his stress tensor, see \citet[pg. 270--273]{whittaker1}; \citet[pg. 168--170, 410--411]{darrigol}.}  This is unfortunate.  The tensor $\sigmaT_{\!f}$ is a momentum flux density tensor, not a stress tensor.\footnote{This observation in no way impugns the use of the Maxwell stress tensor as the spatial-spatial part of the four-dimensional electromagnetic stress-energy tensor which characterizes the flow of the electromagnetic field's energy and momentum.  It does, however, suggest that of the different names sometimes used for the stress-energy tensor, ``energy-momentum tensor'' may be the most perspicuous.}  The tensor $\sigmaT_{\!f}$ gives the flow of momentum per unit area (when dotted with the unit normal $\hat{n}$) and thus is the correct tensor to use in an Eulerian force law \eqref{integralform} where we are concerned with determining the flow of momentum in and out of a fixed volume.  But, the force per unit area exerted across the surface is different, given instead by $\sigmaT_{\!f} + \rho_f\left(\vec{v}_f\otimes\vec{v}_f\right)$.  It is this tensor which is used to calculate surface forces in the Lagrangian force law \eqref{finalforcelaw}.

To see the difference, consider a plane wave in vacuum.  The momentum flux density tensor, $\sigmaT_{\!f}$, will be non-zero as there is a flow of mass in the direction of wave propagation.  However, the true stress tensor, $\sigmaT_{\!f} + \rho_f\left(\vec{v}_f\otimes\vec{v}_f\right)$, will be zero since the field's mass is everywhere flowing steadily without impediment or assistance from the surrounding field.\footnote{In more detail: Consider a linearly polarized plane wave propagating in the $x$-direction as described at a moment by $\vec{E}(\vec{x})=E_0 \cos(k x) \hat{y}$ and $\vec{B}(\vec{x})=E_0 \cos(k x) \hat{z}$.  The momentum flux density tensor can be calculated via \eqref{maxwellstresstensor} yielding a single non-zero component, $\sigmaT_{\!f}=-\frac{1}{4\pi}E_0^2\cos^2(k x) \ \hat{x}\otimes\hat{x}$, describing a flow of momentum in the direction of the electromagnetic wave's propagation.  The rate of change of the field's momentum density in \eqref{momentumconservation} is the divergence of this tensor, $\vec{\nabla}\cdot \sigmaT_{\!f}=\frac{1}{2\pi}E_0^2\sin(2 k x) \hat{x}$.  Because $\rho_f\left(\vec{v}_f\otimes\vec{v}_f\right)$ as calculated using \eqref{massdensityfield} and \eqref{fieldvelocity} is $\frac{1}{4\pi}E_0^2\cos^2(k x) \ \hat{x}\otimes\hat{x}$, the true stress tensor $\sigmaT_{\!m} + \rho_m\left(\vec{v}_m\otimes\vec{v}_m\right)$ is zero.\label{planewavedetail}}

We have seen that the electromagnetic field exerts surface forces.  It is important to keep in mind that these are forces from the field on the field.  The electromagnetic field does not exert surface forces on matter.  It only exerts a body force with density $\vec{f}$.  Still, one can calculate the force on the matter within a volume by integrating the Maxwell stress tensor over the surface enclosing that volume in special cases---when the field momentum in the volume is unchanging, $\iiint \frac{\partial}{\partial t}(\rho_f \vec{v}_f) dV=0$.  Similarly, one can calculate the force on the matter within a volume by integrating the true stress tensor when there is no net force on the field in that volume, $\frac{D}{Dt} \iiint \rho_f \vec{v}_f \:dV=0$.

The Lagrangian force law \eqref{finalforcelaw} can be expressed in differential form by expanding only the second material derivative in \eqref{stepxxxx}, using \eqref{emcontinuity}, equating the expression thus arrived at with \eqref{almostthere}, and dropping the volume integral,
\begin{equation}
\rho_f \frac{D\vec{v}_f}{Dt}= \del \cdot \left[\sigmaT_{\!f} + \rho_f\left(\vec{v}_f\otimes\vec{v}_f\right)\right]-\vec{f}+\frac{\vec{v}_f}{c^2}\left(\vec{f}\cdot\vec{v}^{\,q}_m\right)\ .
\label{finalforcelawdiffform}
\end{equation}
This form of the force law resembles \eqref{relativisticforcealternative} where force is related to mass times acceleration (the force per unit volume on the field being $-\vec{f}\:$).  This further illustrates that the field's mass, $\rho_f$, is truly an inertial mass quantifying the field's resistance to acceleration, $\vec{a}=\frac{D\vec{v}_f}{Dt}$.

Although the analysis thus far has focused on the electromagnetic field, the idea that forces act on fields is quite general.  For any field with well-defined energy and momentum densities, we can introduce a relativistic mass density which is proportional to the energy density and a velocity field which is the momentum density divided by the relativistic mass density.  If that field obeys laws of conservation of energy and conservation of momentum in its interaction with matter (and other fields), then equations of the form of \eqref{emcontinuity} and \eqref{momentumconservationfield} will hold.  From \eqref{momentumconservationfield}, \eqref{finalforcelaw} can be derived.

\section{Field as Fluid}\label{fieldasfluid}

In the preceding sections I have argued that Newton's third law holds in electromagnetism by showing that the electromagnetic field acts very much like a relativistic fluid.  Here are three questions that might come to mind upon encountering this analogy:  First, can we take these fluid equations as among the fundamental equations of electromagnetism, supplanting Maxwell's equations?  Second, can we understand the field's mass density and velocity as describing properties of a large number of discrete particles at a coarse-grained level in the way that mass density and velocity are understood for ordinary fluids?  Third, why has the discussion thus far focused so heavily on relativistic mass instead of proper mass?  One section will be devoted to each of these three questions.  These sections can be read in any order or skipped depending on the reader's curiosities.

The thorough and consistent similarity between the equations with the $f$ (field) subscripts and the equations with the $m$ (matter) subscripts suggests that perhaps it is possible to give an alternative ontology for electromagnetism.\footnote{As was noted at the end of section \ref{forcelawforfield}, other non-electromagnetic fields will generally also resemble fluids in this way.  One might take this as evidence that the electromagnetic field is not particularly fluid-like or as evidence that we may be able to reformulate all field theories as fluid theories.}  
Electromagnetism is commonly understood to be a theory describing the interaction between matter and field.  The electromagnetic field appears to be a quite different sort of thing from matter, described by field variables (such as $\vec{E}$ and $\vec{B}$, the four-dimensional Faraday tensor field $F^{\mu\nu}$, or the electromagnetic four-potential $A^{\mu}$) evolving in accordance with field equations.  But, the discussion in the previous sections suggests that it may be possible to treat the field as just more matter, the state of which can be represented by fluid variables (including the mass density $\rho_f$ and velocity field $\vec{v}_f$) evolving in accordance with fluid equations (such as \eqref{emcontinuity} and \eqref{momentumconservationfield}, in the Eulerian picture).  Maxwell's equations, the Lorentz force law, and the use of electric and magnetic fields could then be interpreted\footnote{Evaluating the merits of such a reinterpretation may involve further comparison of field to fluid (e.g., asking whether the electromagnetic field is composed of particles, as in section \ref{numberdensitysec}).} as non-fundamental ways of describing the evolution of this fluid---a cloak which hides the true nature of the electromagnetic field.\footnote{Taking the electromagnetic field to truly be a fluid would, in a sense, mark a return to the idea of a luminiferous aether---a substance filling all of space within which electromagnetic waves propagate.  However, on this picture electromagnetic waves are really more like gusts of wind (flows of mass) than sound waves (propagating disturbances in a largely stationary medium).  Unlike the old aether theories, this proposal is not intended to recover an absolute notion of rest or simultaneity.  The electromagnetic field would be interpreted as a \emph{relativistic} fluid, something entirely in harmony with the tenets of special relativity.}  In this section I will explain briefly why I think such a reformulation would be valuable, why the preceding discussion does not immediately yield such a reformulation, and how the electromagnetic field differs from an ordinary fluid.

Although there may not be experimentally detectable differences between taking the electromagnetic field to be fundamentally a field governed by Maxwell's equations or a kind of matter governed by fluid equations, I still think it would be worthwhile to develop such a reformulation of the theory.  One of Einstein's primary motivations in developing his special theory of relativity was to unify the electric and magnetic fields.  The Faraday tensor field is often taken to be that unification.\footnote{Another candidate for this unification is the complex-valued vector field $\vec{F}=\vec{E}+i\vec{B}$ (\citealp{good1957}; \citealp[section 16.II.A]{dresden}; \citealp{bialynicki1996}; \citealp{holland2005}).}  But, fields are somewhat exotic and tensor fields are more so.  If the electromagnetic field can instead be understood as a fluid, then in unifying the electric and magnetic fields we move from the more arcane to the more familiar, not vice versa.  Apart from this more metaphysical motivation, there is reason to seek such a reformulation for the sake of doing physics. It is generally useful to have alternative formulations of physical theories at hand when we are solving problems or developing new theories \citep[chapter 7]{feynman}.\footnote{Hertz wrote that although ``we should in no wise confuse the simple and homely figure [of Maxwell's theory], as it is presented to us by nature, with the gay garment with which we use to clothe it'' (which is a matter of interpretation and cannot be determined by experiment), such inventive tailoring is capable of ``aiding our powers of imagination'' (\citealp[pg. 28]{hertz}; discussed in \citealp[pg. 212--215]{hesse}).}  As a third motivation, such a reformulation could improve our understanding of symmetries.  For example, once you replace the field with a fluid, it should be easier to show that the electromagnetism is truly time reversal invariant and invariant under Lorentz transformations.  \citet[chapter 1]{albert} has pointed out that time-reversing a sequence of states (in which the particle locations and field values are fully specified) in the most obvious way (without flipping the sign of $\vec{E}$ or $\vec{B}$) does not generally take law-abiding sequences to law-abiding sequences.  But, a time-reversed history of the fluid would clearly involve flipping $\vec{v}_f$ and this might explain the flip in $\vec{B}$ required for the theory to be time-reversal invariant (looking at \eqref{poyntingvector} and \eqref{fieldvelocity}, this seems plausible).  Similarly, the way $\vec{E}$ and $\vec{B}$ transform under Lorentz transformations is non-trivial and often derived from the assumption that the theory is Lorentz invariant (e.g., \citealp[section 12.3.2]{griffiths}).  One might hope to derive these transformation rules from less contestable transformation laws for the fluid.\footnote{This potential advantage in understanding symmetries using a fluid ontology is analogous to the advantage of a fluid ontology for quantum mechanics in understanding symmetries discussed in \citet[section 12]{sebens2015}.}

If we are to replace the fundamental equations of relativistic electromagnetism with fluid equations, like \eqref{emcontinuity} and \eqref{momentumconservationfield}, we must introduce additional variables to describe the state of this fluid beyond those we've seen thus far.  The Maxwell stress tensor---which is needed to calculate the dynamics of the fluid using \eqref{momentumconservationfield} or \eqref{finalforcelaw}---cannot be determined from the mass density and the velocity alone. (Too see this, imagine rotating the electric and magnetic field together about $\vec{v}_f$.  This will generally change $\sigmaT_{\!f}$ but cannot affect $\rho_f$ or $\vec{v}_f$.)  Further, even if we have $\rho_f$, $\vec{v}_f$, and $\sigmaT_{\!f}$ at our disposal (which contain the same information as the four-dimensional stress-energy tensor of the electromagnetic field), we will not know enough about the field to calculate the force on matter.  (Consider swapping the electric and magnetic fields and multiplying one of the fields by $-1$.  By inspection of \eqref{maxwellstresstensor}, \eqref{massdensityfield}, and \eqref{fieldvelocity} it is clear that this will not effect $\rho_f$, $\vec{v}_f$, or $\sigmaT_{\!f}$.  But, it will in general change the forces felt by charged bodies.\footnote{If you knew the rates of change of $\rho_f$ and $\vec{v}_f$, you could calculate the Lorentz force by \eqref{momentumconservationfield}.  But, then this equation would no longer be able to tell you how the field evolves.})  Thus, there must be additional degrees of freedom characterizing the state of the fluid beyond just $\rho_f$ and $\vec{v}_f$ and they must include information not present in $\sigmaT_{\!f}$.  To make this proposal for new fundamental laws work, one would have to identify appropriate variables to represent these additional degrees of freedom, physically interpret them, and find additional equations to govern their time evolution.\footnote{If we think of the field as a fluid of photons, it may be possible to understand the additional degrees of freedom as describing---in aggregate---the spins/polarizations of the photons as this information does not seem to be captured by the mass density or velocity of the field.}  \citet{bialynicki1997, bialynicki2003} have devised a way of introducing such additional variables, though the resulting equations are not simple and the physical interpretation of the new variables is not entirely clear.  \citet{holland2005} has proposed an alternative way of introducing fluid variables to describe the electromagnetic field.

As was mentioned in section \ref{fmfv}, in describing the matter as a ``fluid'' I have been speaking somewhat loosely.  The use of a mass density and velocity field obeying equations like \eqref{momentumconservationfluid}, \eqref{emcontinuity}, and \eqref{finalforcelawmatter} to describe a substance does not determine whether it is fluid or solid.  Similarly, in asking here whether we might be able to understand the field as a fluid, I am really asking whether we can understand it using the framework of continuum mechanics (which encompasses both solid and fluid mechanics).  It is important to recognize that there are important ways in which the electromagnetic field will not resemble a true fluid.

Consider a constant electric field between two oppositely charged parallel plates separated from one another by a gap in the $x$-direction, $\vec{E}=E \hat{x}$.  Because there is no magnetic field, the field velocity is zero.  The momentum flux density tensor is
\begin{equation}
\sigmaT_{\!f}=\left(\begin{matrix}
  \frac{E^2}{8 \pi} & 0 & 0 \\
  0 &  \frac{-E^2}{8 \pi} & 0 \\
  0 & 0 &  \frac{-E^2}{8 \pi}
 \end{matrix}\right)\ ,
\end{equation}
by \eqref{maxwellstresstensor}.  As $\vec{v}_f=0$, the momentum flux density tensor, $\sigmaT_{\!f}$, is equal to the stress tensor for the field, $\sigmaT_{\!f} + \rho_f\left(\vec{v}_f\otimes\vec{v}_f\right)$.  The force per unit area with normal $\hat{n}$ is thus $\sigmaT_{\!f}\cdot\hat{n}$.  If $\hat{n}$ points in the $y$ or $z$-direction, the force is compressive.  But, if $\hat{n}$ points in the $x$-direction the force has the opposite sign.  The electromagnetic field is sustaining a tensile force (pulling the two plates towards one another).  Solids can sustain such tensile forces but fluids cannot \citep[pg. 70]{mihalas}.  In this way, the electromagnetic field more closely resembles a solid than a fluid.  To see another disanalogy between the electromagnetic field and a true fluid, take $\hat{n}$ to point halfway between the $x$ and $y$-directions.  In this case, the force per unit area, $\sigmaT_{\!f}\cdot\hat{n}$, will not be parallel to $\hat{n}$, indicating the presence of shearing forces.  An ordinary fluid cannot sustain such shearing forces when at rest (\citealp[pg. 12--14]{batchelor}; \citealp[pg. 105]{gurtin}; \citealp[pg. 70]{mihalas}).

\section{Photons and the Classical Electromagnetic Field}\label{numberdensitysec}

Relativistic quantum field theories can either be understood as quantum theories of relativistic fields or relativistic theories of quantum particles.  In particular, the photons which mediate electromagnetic interactions between charged particles in quantum electrodynamics may be understood either as field or particle.  There is debate about whether particle or field is more fundamental in quantum field theory, but to some degree the theory can be interpreted either way.  What I would like to explore in this section is the extent to which we are already able to treat photons as either particle or field at the classical level.

The mass density and velocity field of a relativistic fluid, $\rho_m$ and $\vec{v}_m$, describe at a coarse-grained level the properties of the particles which compose the fluid.  Can we similarly understand the mass density and velocity field of the electromagnetic field as describing properties of the photons which compose the field?  Photons have no proper mass and always travel at the speed of light. These features make them unusual but not unsuitable candidates for a fluid-level description.\footnote{Also, the mean free path for photons is typically large (light beams generally pass through one another undisturbed).  So, photons at some location $(\vec{x},t)$ are likely to have velocities far from the mean velocity $\vec{v}_f(\vec{x},t)$ and not to be carried along with the flow. (See \citealp[chapter 1]{shu2}.)}  Even if each particle is traveling at $c$, $|\vec{v}_f|$ may still be less than the speed of light as $\vec{v}_f$ describes the net flow of relativistic mass and not the actual velocity of any individual particle.  Further, (as was discussed when the field velocity was introduced) $|\vec{v}_f|$ cannot be greater than $c$, which we may view as explained by the fact that the field is made of photons each traveling at $c$.  There is no problem with the field having a non-trivial mass density even if its constituent particles are massless.  Particles without proper mass can still have relativistic mass and collections of particles without proper mass will generally together have proper mass---proportional to the energy they posses in their collective center of momentum frame (see section \ref{propermassdensitysec}).

Thus far I have focused primarily on the relativistic mass densities of fluids and fields.  But, a relativistic fluid also has a number density and if the electromagnetic field is composed of particles it seems like it should have one too.  Because photons are quantum particles, we must look to quantum physics to find a number density for the electromagnetic field.  The possibility of finding a well-defined number-density is threatened by the fact that quantum mechanically there can be a superposition of different number densities (see \citealp[pg. 545]{holland}).

It is possible to give a ``classical'' expression for the photon number density without considering the full complexities of quantum physics. Consider the electromagnetic field in vacuum.  A natural strategy for finding the photon number density would be to divide the energy density of the field, $\rho^{\,\mathcal{E}}_f(\vec{x})=\frac{1}{8\pi}(|\vec{E}(\vec{x})|^2+|\vec{B}(\vec{x})|^2)$, by the energy per photon.  From quantum physics, we know that the energy of a photon with wave vector $\vec{k}$ is $\mathcal{E}= |\vec{k}| \hbar c$.  Because of the dependence on $\vec{k}$, it is hard to execute that natural strategy.  But, it is possible to divide the energy density by the energy per photon to arrive at a number density if we work in $\vec{k}$-space.  Integrating this number density over $\vec{k}$-space gives the total number of photons,
\begin{equation}
N=\iiint{ \frac{|\vec{\widetilde{E}}(\vec{k})|^2+|\vec{\widetilde{B}}(\vec{k})|^2}{8 \pi  |\vec{k}| \hbar c} d^3 k}
\ ,
\label{fieldnumber}
\end{equation}
where $\vec{\widetilde{E}}(\vec{k})$ and $\vec{\widetilde{B}}(\vec{k})$ are the Fourier transformed electric and magnetic fields,\footnote{Because the transformed fields are complex-valued, $|\vec{\widetilde{E}}(\vec{k})|^2$ is shorthand for $\vec{\widetilde{E}}^*\!\!(\vec{k})\cdot\vec{\widetilde{E}}(\vec{k})$.}
\begin{align}
\vec{\widetilde{E}}(\vec{k})&=\frac{1}{(2\pi)^{3/2}}\iiint{ \vec{E}(\vec{x}) e^{- i \vec{k} \cdot \vec{x}} d^3 x}
\nonumber
\\
\vec{\widetilde{B}}(\vec{k})&=\frac{1}{(2\pi)^{3/2}}\iiint{ \vec{B}(\vec{x}) e^{- i \vec{k} \cdot \vec{x}} d^3 x}
\ .
\label{ftransforms}
\end{align}
If you expand $\vec{\widetilde{E}}(\vec{k})$ and $\vec{\widetilde{B}}(\vec{k})$ in \eqref{fieldnumber} using \eqref{ftransforms} and perform the integral over $\vec{k}$, \eqref{fieldnumber} becomes
\begin{equation}
N=\frac{1}{16 \pi^3 \hbar c}\iiint{\!\!\!\iiint{ \frac{\vec{E}(\vec{x})\cdot \vec{E}(\vec{y})+\vec{B}(\vec{x})\cdot \vec{B}(\vec{y})}{|\vec{x}-\vec{y}|^2} d^3 x\: d^3 y}}
\ .
\end{equation}
(This method of counting photons was proposed by \citealp{zeldovich}; see also \citealp[pg. 318]{bialynicki1996}; \citealp{avron}.)  One can generate a number density by simply dropping one of the spatial integrals,
\begin{align}
\rho^{\,N}_f\!(\vec{x})&=\frac{1}{16 \pi^3  \hbar c}\iiint{ \frac{\vec{E}(\vec{x})\cdot \vec{E}(\vec{y})+\vec{B}(\vec{x})\cdot \vec{B}(\vec{y})}{|\vec{x}-\vec{y}|^2} d^3 y}
\nonumber
\\
&=\frac{1}{16 \pi^3  \hbar c}\left[\vec{E}(\vec{x})\cdot\iiint{\frac{\vec{E}(\vec{x}+\vec{r})}{r^2} d^3 r}+\vec{B}(\vec{x})\cdot\iiint{\frac{\vec{B}(\vec{x}+\vec{r})}{r^2} d^3 r}\right]
\ ,
\label{fieldnumberdensity}
\end{align}
with $\vec{r}$ defined as $\vec{y}-\vec{x}$. It is odd that the density of photons at one point depends on what's happening everywhere else (through the spatial integral), but not obviously unacceptable---figuring out what kind of photons there are at a certain point may require examining the way the field behaves nearby (nearby field values are most important as the factor of $|\vec{x}-\vec{y}|^2$ in the denominator suppresses contributions from distant locations).

To check that \eqref{fieldnumberdensity} is at least a minimally sensible expression, consider the simple case of a linearly polarized plane wave propagating in the $x$-direction with $\vec{E}(\vec{x})=E_0 \cos(k x) \hat{y}$ and $\vec{B}(\vec{x})=E_0 \cos(k x) \hat{z}$.  The energy density of the plane wave is $\frac{E_0^2}{4\pi}\cos^2(k x)$ and the energy per photon is $k \hbar c$.  As the reader can confirm, the photon density derived using \eqref{fieldnumberdensity} is $\frac{E_0^2 \cos^2(k x)}{4\pi k \hbar c}$, exactly as it should be.

\section{Proper Mass Density}\label{propermassdensitysec}

In attributing fluid-like properties to the electromagnetic field, we have not yet given it a proper mass density.  The reason for this is that there are a number of oddities involved in defining the proper mass density for a relativistic fluid.  It is no more difficult to define a proper mass density for the electromagnetic field, but the oddities afflicting the fluid case are inherited.

There are five densities for a fluid that have come up in our discussion thus far: charge density, energy density, relativistic mass density, proper mass density, and particle number density.  Corresponding to these five densities are five velocity fields which describe their flows: $\vec{v}^{\,q}_m$, $\vec{v}^{\,\mathcal{E}}_m$, $\vec{v}^{\,m_r}_m$, $\vec{v}^{\,m_0}_m$, and $\vec{v}^{\,N}_m$.  Are any of these five velocity fields equal to one another?  For the most part, no.  The only equality that holds in general is between the velocities describing energy and relativistic mass flow (since energy is directly proportional to relativistic mass).  This velocity, $\vec{v}^{\,\mathcal{E}}_m=\vec{v}^{\,m_r}_m$, is the velocity we have focused on thus far and denoted $\vec{v}_m$ (without a superscript).  This velocity is generally not equal to the velocity describing the flow of particle number, $\vec{v}^{\,N}_m$, because sometimes energy (and thus relativistic mass) flows as a result of something other than the bulk motion of particles in the fluid, e.g., when there is heat flux (see \citealp[pg. 505]{landaulifshitzfluids}). If we add the assumption that the particles which compose the fluid have equal charge $q$, then the charge density is $q$ times the number density and $\vec{v}^{\,q}_m$ is equal to $\vec{v}^{\,N}_m$ (but because $\vec{v}_m \neq \vec{v}^{\,N}_m$, $\vec{v}^{\,q}_m$ will generally differ from $\vec{v}_m$, as was mentioned in section \ref{fmfv}).  If the particles have equal proper mass $m_0$ and we assume (as an idealization) that all of the particles near $(\vec{x},t)$ can be treated (contra footnote \ref{average}) as moving with the same velocity $\vec{v}^{\,N}_m(\vec{x},t)$, then the proper mass density is simply $m_0$ times the number density and, because of this proportionality, the velocity describing the flow of proper mass, $\vec{v}^{\,m_0}_m$, is equal to the velocity describing the flow of particle number, $\vec{v}^{\,N}_m$ (see \citealp[equation 39.5]{mihalas}).  Relaxing this idealization, the proper mass density can be defined in two different ways, each of which has undesirable features.

The first method is to tie proper mass density closely to number density, insisting that the proper mass density is the particle proper mass $m_0$ times the number density and that $\vec{v}^{\,N}_m$ is, as before, equal to $\vec{v}^{\,m_0}_m$.  As number density is locally conserved and transforms trivially under Lorentz transformations (picking up a factor of $\gamma$ since it is a density and length contraction shrinks the volumes over which it is integrated, \citealp[eq. 39.4]{mihalas}), proper mass density so defined will inherit these properties.  But, with this definition there is little connection between proper mass density and energy density (or relativistic mass density).  To see why this is so, consider a simple case in which there is no flow of particle number and no flow of energy, $\vec{v}^{\,N}_m=\vec{v}^{\,\mathcal{E}}_m=0$.  Because the proper mass of a system in its rest frame is equal to the system's energy divided by $c^2$, one might hope that in this state of rest the fluid's energy density is equal to the proper mass density times $c^2$.  With this definition, that will not be so.  The proper mass density defined as $\rho^{\,m_0}_m=m_0\times\rho^{\,N}_m$ does not account for contributions to the energy in this frame other than the rest mass energies of the particles, such as the contribution from the heat of the fluid (the energy due the random motion of particles about their mean velocity).

The second method for defining the proper mass density of a fluid ties it more closely to relativistic mass density and energy density.  The proper mass density is defined in terms of the energy density, $\rho^{\,\mathcal{E}}_m=\rho_m c^2$, and the momentum density, $\rho_m \vec{v}_m$, by $\rho_m^2 c^2-\rho_m^2 |\vec{v}_m|^2=(\rho^{\, m_0}_m)^2 c^2$ (taking the relativistic relation between energy, momentum, and proper mass $(\mathcal{E}/c)^2-|\vec{p}|^2=m_0^2 c^2$, as applied to the densities of these quantities, to be definitional of proper mass density).  From this equation it follows that $\rho^{\, m_0}_m=\frac{\rho_m}{\gamma}=\frac{\rho^{\,\mathcal{E}}_m}{\gamma c^2} $ where the velocity that appears in $\gamma$ is the velocity that describes the flow of relativistic mass, $\vec{v}_m$.  If we consider the fluid at a point from a frame where the momentum density at that point is zero, the proper mass density will simply be given by the energy density at that point divided by $c^2$.  This definition resolves the problem raised for the previous definition since all contributions to the energy are included ($\rho^{\, m_0}_m$ will thus generally be greater than $m_0\times\rho^{\,N}_m$).  If we idealize away the effect of heat by assuming that all of the particles near $(\vec{x},t)$ have velocity $\vec{v}^{\,N}_m(\vec{x},t)$ and assume that there are no other contributions to the energy that need to be included beyond the rest mass and kinetic energies of the particles, then $\vec{v}^{\,N}_m=\vec{v}_m$, the energy density is $\rho^{\,\mathcal{E}}_m= m_0 \gamma c^2 \times\rho^{\,N}_m$, and the rest mass density is exactly $\rho^{\,m_0}_m= m_0 \times \rho^{\,N}_m$ (the two alternative definitions agree).  On this second definition, local conservation of proper mass cannot be derived straightforwardly from either conservation of particle number or conservation of energy.  This makes it difficult to find an appropriate velocity $\vec{v}^{\,m_0}_m$ to describe the flow of proper mass.

On neither of the two proposed definitions will integrating the proper mass density of the fluid over the volume occupied by the fluid give the proper mass of the fluid as a whole.  This is because proper mass is not additive.  Consider the collection of particles that compose a gas as described by their positions and velocities, not yet by densities and velocity fields.  Summing the proper masses of the particles will not give the proper mass of the gas.  The proper mass of the gas is found by dividing the energy of the gas by $c^2$ in the frame where the net momentum of the gas is zero---the rest frame of the gas (see \citealp[chapter 8]{lange2001, lange}).  Similarly, the proper mass of a fluid described using densities and velocity fields can be determined by integrating the fluid's energy density in the frame in which the net momentum of the fluid is zero and dividing by $c^2$.

As with a relativistic fluid, the field's proper mass density can be defined in two ways.  If we define it as the mass of each particle times the number density and take the field to be composed of massless photons, then the proper mass density is everywhere and always zero.  Alternatively, we can derive a proper mass density using the relation $(\mathcal{E}/c)^2-|\vec{p}|^2=m_0^2 c^2$ as applied to the energy, momentum, and proper mass densities of the field (see \citealp[pg. 244]{lange}):
\begin{equation}
\rho^{\,m_0}_f=\frac{1}{4 \pi c^2}\sqrt{\frac{1}{4}\left(E^2-B^2\right)^2+(\vec{E}\cdot\vec{B})^2}\ .
\label{fieldpropermassdensity}
\end{equation}
We saw earlier that the proper mass density of a fluid (defined in this second way) can be more than the particle mass times the number density, $m_0\times\rho^{\,N}_m$, because of other contributions to the energy---e.g., random particle motion a.k.a. heat.  If a fluid is made of massless particles its proper mass must come entirely from such contributions as $m_0=0$.  Note that the proper mass density in \eqref{fieldpropermassdensity} is frame-independent\footnote{See \citet[pg. 244]{lange} alongside \citet[problem 11.14]{jackson}; \citet[section 161]{garg}.} whereas the proper mass density for a relativistic fluid of massive particles (similarly defined) is frame-dependent.  With proper mass density as with speed, only fleet-footed photons manage to appear the same in every frame.\footnote{In \citet{lange} the guiding question is whether electromagnetic interactions are local.  Lange argues that this turns on the question of whether the electromagnetic field is real.  On pg. 247 he summarizes his case that it is:
\begin{quote}
``...we tried to use the ontological status of energy and momentum to support the field's reality ... But this argument ultimately failed when we learned from relativity theory that energy and momentum are frame-dependent, and hence unreal.  Relativity did, however, reveal [proper] mass to be real.  Its ontological status has now come to underwrite the electromagnetic field's reality, since the field has turned out to possess [proper] mass.''
\end{quote}
The analysis in this section challenges that reasoning.  It is true that, as Lange emphasizes, the proper mass density of the field as defined by \eqref{fieldpropermassdensity} is frame-independent.  But, this is a special feature of the electromagnetic field and thus an odd hook on which to hang its reality.  The proper mass density for a relativistic fluid similarly defined will be frame-dependent, and yet we do not doubt whether such fluids are real.  Of course, there is another reason available for taking relativistic fluids to be real: the particles that compose such fluids have frame-invariant proper masses.}

As was the case for a relativistic fluid, the proper mass of the field as a whole is not the integral of the proper mass density.  Instead, it can be determined by integrating the field's energy density in the frame in which the net momentum of the field is zero and dividing by $c^2$.  This method of determining the field's proper mass raises problems.  There may be no rest frame for the field as a whole (as would happen if, for example, the field is that of a plane wave traveling at $c$; see \citealp{rohrlich1970}).  To avoid this problem we can define the proper mass of the field in a particular frame in terms of the field's total energy and momentum in that frame by $(\mathcal{E}/c)^2-|\vec{p}|^2=m_0^2 c^2$.  Because the three-dimensional hyperplane of simultaneity in a moving frame will not agree with the hyperplane of simultaneity in the original frame and in the gaps between these hyperplanes there may be an exchange of energy (and momentum) between the field and matter, the field's proper mass will end up differing from frame to frame.  This can seem problematic when it is put as a lament that the field's energy and momentum do not transform as a four-vector: the inner product of the energy-momentum four-vector with itself, $(\mathcal{E}/c)^2-|\vec{p}|^2=m_0^2 c^2$, is frame-dependent. (If there is no matter present and the field is truly isolated this problem will not arise.\footnote{This can be put another way: the total energy and momentum of the field will transform as a four-vector if there is no matter present because the divergence of the stress-energy tensor is zero everywhere \citep[problem 12.18]{jackson}.})  In response to this second problem, we can either define the field's energy, momentum, and proper mass as relative to a specified hyperplane (\citealp{rohrlich1970, rohrlich26, rohrlich}) or just note that only when we speak of matter and field together will energy and momentum transform as a four-vector and proper mass be frame-invariant (\citealp{griffithsowen}; \citealp{rohrlich26}).

\section{Conclusion}

In electromagnetism, as in Newton's mechanics, action is always equal to reaction.  The force from the electromagnetic field on matter is balanced by an equal and opposite force from matter on the field.  The response of the field to this force can be given in Eulerian \eqref{momentumconservationfield}, \eqref{integralform} or Lagrangian \eqref{finalforcelaw}, \eqref{finalforcelawdiffform} form.  These equations perfectly match those that govern a relativistic fluid.  From examination of the analogy it is clear that the mass of the field plays exactly the same inertial role as the mass of a fluid and also that the Maxwell stress tensor is really a momentum flux density tensor, not a stress tensor.  This perfect match of field and fluid equations suggests that it may be possible to reinterpret the electromagnetic field as a fluid of photons.\\\\

\noindent
\textbf{Acknowledgments}

\noindent Thank you to Craig Callender, Dirk-Andr\'{e} Deckert, Mario Hubert, and Mark Lange for valuable discussions and helpful feedback on drafts of the paper.  Special thanks to Erik Curiel, Dennis Lehmkuhl, and James Weatherall.  Thanks also to the anonymous referees who provided useful comments on the paper.

\bibliography{relEMbibfile}

\end{document}